%% file: amia.tex
\documentclass{amia}
\usepackage{lipsum} 
\usepackage{graphicx}
\usepackage{booktabs}
\usepackage{placeins}
\usepackage{multirow}
\usepackage{colortbl}
\usepackage{xcolor}
\usepackage{algorithm}
\usepackage{algorithmic}
\usepackage{amssymb}
\usepackage{float} 
\usepackage{amsmath}
\usepackage{verbatim}
\usepackage{tabularx}
\usepackage{paralist}
\usepackage{soul}
\usepackage[T1]{fontenc}

\begin{document}

\title{SeizureFormer: A Transformer Model for IEA-Based Seizure Risk Forecasting}

\author{
    Tianning Feng$^1$, Juntong Ni$^1$, 
    Ezequiel Gleichgerrcht, MD, PhD$^2$, Wei Jin, PhD$^1$
}
\institutes{
    $^1$Department of Computer Science, Emory University, Atlanta, GA \\ $^2$Department of Neurology, Emory University, Atlanta, GA
}
\maketitle

\section*{Abstract}
\vspace{-4mm}
We present \textbf{SeizureFormer}, a Transformer-based model for long-term seizure risk forecasting using interictal epileptiform activity (IEA) surrogate biomarkers and long episode (LE) biomarkers from responsive neurostimulation (RNS) systems. Unlike raw scalp EEG-based models, SeizureFormer leverages structured, clinically relevant features and integrates CNN-based patch embedding, multi-head self-attention, and squeeze-and-excitation blocks to model both short-term dynamics and long-term seizure cycles. Tested across five patients and multiple prediction windows (1–14 days), SeizureFormer achieved state-of-the-art performance with \textbf{mean ROC AUC} of \textit{79.44\%} and \textbf{mean PR AUC} of \textit{76.29\%}. Compared to statistical, machine learning, and deep learning baselines, it demonstrates enhanced generalizability and seizure risk forecasting performance under class imbalance. This work supports future clinical integration of interpretable and robust seizure forecasting tools for personalized epilepsy management.

\input{section/intro}
\input{section/related}
\input{section/method.tex}
\input{section/experi.tex}

\input{section/result.tex}
\input{section/conclusion.tex}
\makeatletter
\renewcommand{\@biblabel}[1]{\hfill #1.}
\makeatother

\bibliographystyle{vancouver}
\bibliography{amia}  

\end{document}

%% file: section/intro.tex
\section*{Introduction}
\vspace{-3mm}
Epilepsy is a chronic neurological disorder affecting around 50 million people worldwide, marked by recurrent and unpredictable seizures that significantly disrupt daily life.~\cite{who2023epilepsy, kerr2024seizureforecasting}. While anti-seizure medications (ASM) may control seizures in 60-65\% of cases~\cite{Kwan2010}, a substantial number of patients continue to have seizures and may require other therapeutic interventions, such as destructive surgical procedures or neuromodulation ~\cite{Fisher2014EpilepsyDefinition}. For these patients, accurately forecasting seizures remains a crucial unmet need in epilepsy management. Reliable seizure risk forecasting would enable patients to take preventive measures and enhance clinical decision-making, ultimately improving patient outcomes~\cite{kerr2024seizureforecasting, kuhlmann2021seizureprediction}.

Traditional seizure prediction methods rely on statistical tools, including frequency-domain techniques (e.g., Fourier and wavelet transforms)~\cite{carney2011seizure, ni2025timedistill}, time-domain techniques (e.g., autocorrelation and variance analysis)~\cite{carney2011seizure}, and nonlinear dynamical measures (e.g., Lyapunov exponent)~\cite{Stam2005NonlinearEEG, Acharya2013AutomatedEEG} to detect and classify pre-seizure patterns. While these statistical approaches have provided valuable insights into seizure detection, their ability to perform seizure forecasting is limited due to their dependence on cyclic patterns~\cite{Proix2021}. Additionally, traditional seizure prediction methods primarily rely on scalp EEG data, which is inherently non-stationary and can hinder the model's performance~\cite{rasheed2021ml, yang2022seizure}. Raw scalp EEG signals are highly susceptible to a range of physiological and external influences—including brain state transitions (e.g., sleep vs. wake), cognitive load, medication effects, fatigue, and changes in electrode impedance or placement. These factors can cause significant fluctuations in EEG amplitude, frequency content, and waveform morphology over time~\cite{rasheed2021ml, yang2022seizure}. Although recent deep learning approaches have enabled short-term EEG-based forecasting (1–24 hours) using frequency-domain features~\cite{turn0search1, turn0search2, turn0search4}, the non-stationary nature of scalp EEG signals fundamentally limits their stability and generalizability, making long-term forecasting inherently challenging. 

As an alternative, responsive neurostimulation (RNS) systems extract seizure-related signals such as interictal epileptiform activity (IEA) and long episodes (LE), which have been validated as reliable seizure risk indicators by reflecting sustained EEG abnormalities linked to cortical excitability~\cite{Khambhati2024, Yang2024SeizureForecasting}. These structured features offer a more stable and interpretable basis for seizure risk assessment compared to scalp EEG signals. Leveraging these biomarkers, recent studies have employed statistical models such as Poisson regression and generalized linear models (GLMs) to capture long-term seizure risk trends~\cite{Proix2021}. Machine learning and deep learning methods, including convolutional neural networks (CNNs) for automatic seizure detection and recurrent neural networks (RNNs) for short-term forecasting, have also been explored~\cite{Peterson2023}. Though these existing seizure forecasting approaches have shown promising performance using RNS data, they still face some key limitations:

\begin{compactenum}[\textbullet]
    \item \textit{Limited multi-scale modeling ability:} RNS data exhibits multi-scale dynamics, with seizure risk influenced by rhythmic patterns at weekly, daily, hourly, and sub-second levels~\cite{schroeder2022tracking}. However, existing work often fails to capture these complex temporal dependencies.
    \item \textit{Limited short-term and long-term dependency modeling ability:} 
    RNS data also exhibits long-term dependencies, where seizure risk is influenced not only by recent activity but also by long-term seizure history records. Statistical models (e.g., Poisson regression, GLMs) are effective for long-term seizure risk trends but lack the granularity for short-term forecasting~\cite{Proix2021}. Conversely, explored deep learning models such as RNNs and CNNs excel in short-term seizure detection but fail to integrate long-range seizure periodicities~\cite{Peterson2023}. A unified approach that can jointly model both short-term variability and long-term seizure cycles is needed.
    \item \textit{Overreliance on raw RNS data, leading to high personalization requirements} – Many existing seizure forecasting models heavily depend on raw RNS data, which are highly complex and patient-specific. As a result, these models often require extensive per-patient training to achieve accurate predictions, limiting their generalizability across individuals~\cite{Khambhati2024, Yang2024SeizureForecasting}. Reducing reliance on raw RNS data with high variability and leveraging more structured and stable biomarkers could improve model scalability and clinical applicability. 
\end{compactenum}

In this work, we aim to design a new deep learning model for seizure forecasting that can jointly tackle the above limitations and effectively capture temporal dynamics in RNS data across multiple timescales. In essence, we are faced with three key challenges. \textbf{First, how to extract seizure-relevant features from time-series signals with varying periodicity?} We address this by employing a CNN-based patch embedding module with multiple kernel sizes, allowing the model to capture both short-term fluctuations and long-term cycles from the input signals. \textbf{Second, how to bridge the gap between short-term and long-term forecasting, which are often handled separately in prior work?} We tackle this by introducing a hierarchical Transformer architecture that jointly models local and global temporal dependencies, enabling unified forecasting across different time horizons. \textbf{Third, how to reduce model dependence on raw, highly individualized EEG signals?} Instead of using raw RNS data, we leveraged the RNS system's continuous monitoring of EEG-based hyperexcitable patterns (A/B spike detections and long episodes) to reduce the reliance on raw signal processing. Together, these components form SeizureFormer, a robust solution for generalized, multi-timescale seizure risk forecasting.

Our model demonstrates leading performance across multiple evaluation settings. Extensive experiments show that SeizureFormer achieves the highest \textbf{ROC AUC (79.44\%)} and \textbf{PR AUC (76.29\%)}, significantly outperforming baseline statistical and recurrent models. Furthermore, our approach extends seizure forecasting beyond conventional short-term windows, enabling \textbf{multi-day (1-14 days) risk estimation}, which is critical for clinical decision-making and patient management. Our contributions can be summarized as follows:
\begin{compactenum}[\textbullet]
    \item We develop the first Transformer-based model for seizure risk forecasting, SeizureFormer, capable of capturing both short-term fluctuations and long-horizon seizure cycles. This enables accurate seizure forecasting across different time scales up to 14 days ahead, addressing the limitations of prior approaches.
    \item Our model relies on A+B spike patterns and Long Episodes (LE) as seizure risk proxy, reducing dependence on large volumes of raw EEG data while maintaining predictive robustness and generalizability.
    \item Extensive experiments demonstrate that SeizureFormer outperforms statistical and deep learning baselines in both short-term and long-term seizure forecasting, achieving superior predictive accuracy and clinical applicability.
    
\end{compactenum}

%% file: section/related.tex
\section*{Related Works}
\vspace{-3mm}

\underline{\textit{EEG-based Seizure Risk Forecasting}}:
Traditional seizure forecasting models rely on scalp EEG data, using statistical and deep learning methods to estimate seizure risk. Statistical approaches such as Poisson regression, GLMs, and logistic regression model seizure probability through probabilistic frameworks~\cite{epilepsia2022}. Proix et al.~\cite{Proix2021} showed that multi-day seizure cycles could be captured using Poisson regression on long-term iEEG, achieving above-chance forecasting in $\sim$66\% of patients. While effective for modeling cyclic trends, these methods struggle with the fine-grained temporal resolution required for real-time forecasting. To address this, deep learning models like RNNs, GRUs, and LSTMs have been applied to capture sequential dependencies in EEG signals~\cite{thodoroff2016deep, raghu2019deep}. Though they improve short-term prediction, they face limitations in modeling long-range patterns and generalizing across patients~\cite{Andrzejak2023}, partly due to the variability of raw EEG data. Recent studies have attempted to incorporate multi-day periodicity and additional physiological biomarkers such as heart rate (HR) and skin conductance (SC)~\cite{Andrzejak2023}. However, few approaches integrate both short-term fluctuations and long-term seizure cycles, limiting their clinical applicability.

\underline{\textit{RNS-Based Seizure Risk Forecasting}}:
The RNS system, compared to raw scalp EEG, provides more stable and temporally correlated biomarkers such as interictal epileptiform activity (IEA) and long episodes (LE), making it a reliable source for seizure-related forecasting. Several studies have explored this potential: Yueqiu et al.\cite{Yueqiu2021} demonstrated the feasibility of using statistical and classical machine learning methods, including SVM and random forests, to forecast seizure frequency from long-term RNS recordings, capturing meaningful patient-specific temporal patterns with AUC ranging from 70\% to 89\%. Deep learning approaches have also been applied; for example, Constantino et al.\cite{Constantino2021} and Peterson et al.~\cite{Peterson2023} utilized CNN-based models for seizure detection and onset prediction, achieving high accuracy on short-term tasks. While these studies show strong potential for leveraging structured RNS data, they still face limitations: most models heavily rely on patient-specific features and require extensive per-patient training due to the raw, individualized nature of RNS signals.

\underline{\textit{Long-Term Seizure Risk Forecasting}}:
Recent work has extended seizure prediction from short-term to long-term horizons using RNS data. For example, Seizure Forecasting with Ultra Long-Term EEG Signals~\cite{clinicalneurophysiology2024} showed that Poisson regression with SVMs achieves AUCs above 70\% for up to 6-day forecasts, highlighting the potential of cyclic RNS features. However, most models target either short- or long-term patterns in isolation, and few explore learning both simultaneously. To address this, we propose a Transformer-based model that captures short-term fluctuations and long-range dependencies via multi-head self-attention. Compared to GRUs and LSTMs, Transformers mitigate vanishing gradients and better model multi-day seizure cycles. We also benchmark statistical, classical ML, and deep learning baselines, showing that our approach achieves superior performance in capturing seizure dynamics across time scales.

%% file: section/method.tex
\section*{Preliminary}
\vspace{-3mm}

In this section, we first present a general problem formulation for seizure risk forecasting. We then describe how structured features extracted from the RNS data are utilized to support the forecasting process.

\underline{\textit{Problem Formulation}}:
We formulate seizure risk forecasting as a binary classification task, where the objective is to predict whether a given day falls into a high-risk seizure period based on historical biomarkers of brain excitability.  Given an input time series:
$\mathbf{X} = [\mathbf{x}_{t-n}, \mathbf{x}_{t-n+1}, ..., \mathbf{x}_t] \in \mathbb{R}^{n \times d}$, where \(X_t\) represents the extracted feature set at time \(t\), \(d\) is the feature dimension (i.e., the number of input variables), and \(n\) is the sequence length, defining the number of past time steps \(\{\mathbf{x}_{t-n}, \mathbf{x}_{t-n+1}, ..., \mathbf{x}_t\}\)
 used as input. The goal is to predict whether the next \(h\) days belong to a high-risk seizure period:
$y_h = f(\mathbf{x}_{t-n}, \mathbf{x}_{t-n+1}, ..., \mathbf{x}_t)$
where \(y_h \in \{0,1\}\) is a binary label determined by the presence of long episodes.

\underline{\textit{RNS-Based Feature and Label Seleciton}}: 
Due to the non-stationary nature of raw scalp EEG, we chose to use data collected by RNS system. While raw RNS data can still be varying among different patients and across time, some clinically relevant biomarkers can be extracted from raw RNS data to represent seizure proxy in a more stable way. 
To capture these clinically relevant dynamics, we extract two structured biomarkers from the RNS system:

\begin{compactenum}[\textbullet]
\item \textit{A+B Patterns (IEA Surrogate Biomarkers)}: The RNS system continuously monitors electrocorticographic activity and detects patient-specific epileptiform discharges, including spikes, sharp waves, and rhythmic bursts. Each implanted device records activity from two channels (depth electrodes or subdural strips), detecting two distinct epileptiform activity patterns (i.e. Pattern A and Pattern B) for each channel. These patterns provide a structured time-series representation of cortical excitability, making them a more stable biomarker for seizure forecasting compared to raw EEG. For each device with two channels, we extract Pattern A and B from both channels to get two combined features: \textit{Pattern A + B on Channel 1} and \textit{Pattern A + B on Channel 2}. These serve as the input features \(\mathbf{x}_t \in \mathbb{R}^2\), and the full input sequence \(\mathbf{X} = [\mathbf{x}_{t-n}, \dots, \mathbf{x}_t] \in \mathbb{R}^{n \times 2}\) is used to predict the seizure risk label \(y_h\).

\item \textit{Long Episodes (LEs) as Seizure Risk Indicators}: The RNS system also records Long Episodes (LEs), defined as sustained abnormal electrocorticographic activity exceeding a predefined duration threshold (typically 10–30 seconds). Prior studies have shown that LE occurrences correlate strongly with seizure likelihood, making them an effective proxy for defining high-risk periods~\cite{Khambhati2024}. In our formulation, the binary label \( y_h \in \{0,1\} \) is determined based on LE activity: we assign \( y_h = 1 \) if the cumulative LE count or duration in the forecasting window (e.g., the next \( h \) days) exceeds a patient-specific threshold, indicating a high-risk period; otherwise, \( y_h = 0 \). This aligns with clinical practice, where seizure risk is assessed probabilistically rather than as discrete events.

\end{compactenum}

\textbf{Our Proposed Model - SeizureFormer}

\begin{figure}[htbp!]
\scriptsize
\centering
\includegraphics[width=1\textwidth]{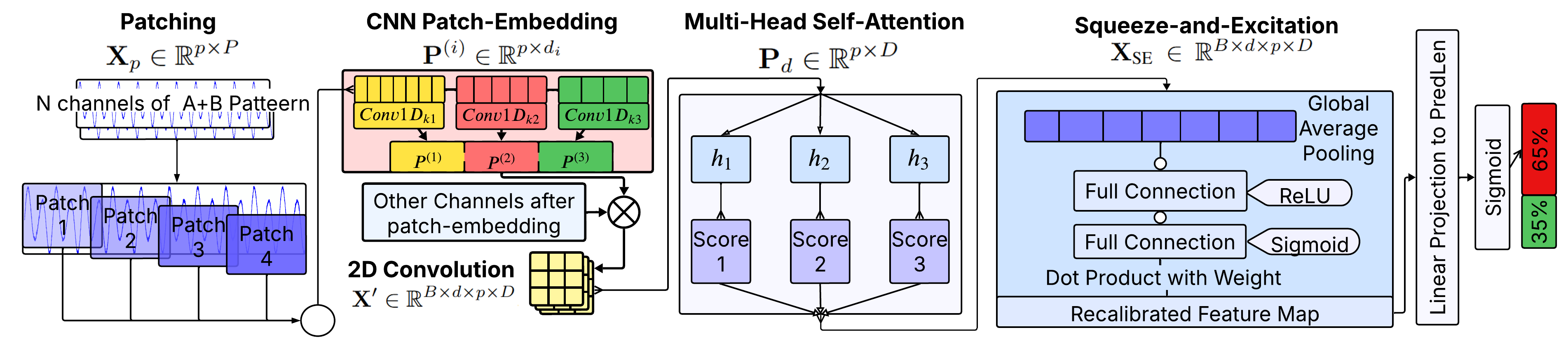}
\caption{Overall framework of SeizureFormer.}
\label{fig:model}
\end{figure}

As shown in Figure~\ref{fig:model}, SeizureFormer consists of five key modules: (1) Patching, (2) CNN Patch-Embedding, (3) Cross-Variable Temporal 2D Conv, (4) Multi-Head Self-Attention, and (5) Squeeze-and-Excitation. In the following, we describe each module in detail.

\noindent
\underline{\textit{Patching}}:  
While there may exist \( d \) A+B pattern channels, each channel is processed independently with the same procedure in the patching module, which we illustrate through a single-channel example. To effectively capture local seizure-relevant patterns, the first step in our model is to segment the input time series into patches. Given a time series \( \mathbf{X} \in \mathbb{R}^{N \times 1} \), where \( N \) is the sequence length and the RNS channel is divided into patches of length \( P \) with stride \( S \):
\begin{equation}
\mathbf{X}_p = \{\mathbf{X}_{t:t+P} \mid t = 1, S, 2S, ..., N - P\}, \quad \mathbf{X}_p \in \mathbb{R}^{p \times P},
\end{equation}
where \( P \) is the \textit{patch length}, and \( S \) is the \textit{stride}, which controls the degree of overlap between consecutive patches. The number of patches is given by \( p = \lfloor (N - P) / S \rfloor + 1 \). If \( S < P \), patches overlap, enabling richer local temporal relationships, whereas \( S = P \) results in non-overlapping patches. This patching process reduces input redundancy while allowing the model to capture localized temporal dependencies.

\noindent
\underline{\textit{CNN Patch-Embedding}}:  
$\mathbf{X}_p$ is then processed through a CNN-based patch embedding module, which extracts temporal features using multiple convolutional filters with varying receptive fields. Specifically, we apply a set of parallel 1D convolutional layers with \( K \) different kernel sizes \( \{k_1, k_2, ..., k_K\} \), each designed to capture temporal patterns at a different resolution. Each convolution outputs the same number of extracted features for every patch, denoted as:
\begin{equation}
\mathbf{P}^{(i)} = \text{Conv1D}_{k_i}(\mathbf{X}_p), \quad \mathbf{P}^{(i)} \in \mathbb{R}^{p \times d_i}, \quad d_1 = d_2 = ... = d_K, \quad i = 1, ..., K,
\end{equation}
where \( \text{Conv1D}_{k_i}(\cdot) \) denotes a 1D convolution with kernel size \( k_i \), and \( d_i \) is the number of features extracted by the \( i \)-th convolution. All convolutions are configured to output the same number of features, i.e., \( d_1 = d_2 = \cdots = d_K \).
The final embedding is obtained by concatenating the outputs along the feature dimension:
\begin{equation}
\mathbf{P} = \text{Concat}\left(\mathbf{P}^{(1)}, \mathbf{P}^{(2)}, ..., \mathbf{P}^{(K)}\right), \quad \mathbf{P} \in \mathbb{R}^{p \times d'},
\end{equation}
where \( d' = \sum_{i=1}^{K} d_i \) is the total embedding dimension. This CNN-based multi-kernel design enables the model to integrate both fine-grained and coarse-grained temporal information from A+B pattern patches, enhancing its ability to detect seizure-relevant dynamics across multiple time scales. To preserve the temporal ordering of patches, we incorporate a learnable position encoding:
\begin{equation}
\mathbf{P}_d = \mathbf{W}_P \mathbf{P} + \mathbf{W}_{\text{pos}}, \quad \mathbf{P}_d \in \mathbb{R}^{p \times D},
\end{equation}
where \( \mathbf{W}_P \in \mathbb{R}^{d' \times D} \) is a trainable projection matrix mapping the patch features to embedding dimension \( D \), and \( \mathbf{W}_{\text{pos}} \in \mathbb{R}^{p \times D} \) is a learnable position encoding that maintains the sequence order. The resulting \( \mathbf{P}_d \) is then used as input to the Transformer encoder.

\noindent
\underline{\textit{Cross-Variable Temporal 2D Conv (CVT-Conv)}}:  
While the patch embedding process is illustrated using a single A+B pattern channel for clarity, in practice, the input contains \( d \) channels. Although patch embeddings capture local temporal patterns within each A+B pattern channel independently, they do not account for potential interactions across channels. To enrich the representation with inter-channel temporal dependencies, we introduce a 2D convolution over the patches from different channels \( (d, p) \).

At a macro level, the embedded tensor after patching and projection is represented as \( \mathbf{X} \in \mathbb{R}^{B \times d \times p \times D} \), where \( B \) is the batch size, \( d \) is the number of A+B pattern channels, \( p \) is the number of temporal patches, and \( D \) is the feature dimension of each patch. This structure forms a 2D grid of patch features for each sample, with spatial axes corresponding to pattern channels and time, and feature vectors residing in the \( D \)-dimensional space.

We apply a 2D convolution over the \( (d, p) \) grid using a learnable kernel \( \mathbf{\mathcal{K}} \in \mathbb{R}^{k_d \times k_p} \), treating the \( D \)-dimensional patch features as individual vectors passed through shared filters. This operation allows each channel-patch pair to aggregate contextual information from its spatial neighborhood across both channel and temporal dimensions, leading to a richer representation before feeding into the self-attention module:
\begin{equation}
\mathbf{X}' = \text{Conv2D}_{\mathbf{\mathcal{K}}}(\mathbf{X}), \quad \mathbf{X}' \in \mathbb{R}^{B \times d \times p \times D}.
\end{equation}
Here, appropriate padding is applied to preserve the spatial and temporal dimensions. The kernel \( \mathbf{\mathcal{K}} \) enables each channel-patch pair to aggregate contextual information from its local neighborhood across both pattern channels and time. The size of \( \mathbf{\mathcal{K}} \) determines how much cross-channel and temporal context is integrated, allowing the model to extract richer inter-channel and sequential features before self-attention. This pre-attention enrichment provides strong inductive bias for capturing local temporal and inter-channel patterns, which complements the global modeling capability of the subsequent self-attention layers.

\noindent
\underline{\textit{Multi-Head Self-Attention (MHSA)}}:  
From the previous module (CVT-Conv), we obtain a 4D tensor \( \mathbf{X}' \in \mathbb{R}^{B \times d \times p \times D} \), where \( B \) is the batch size, \( d \) is the number of A+B pattern channels, \( p \) is the number of temporal patches per channel, and \( D \) is the feature dimension of each patch.

To apply attention independently for each channel and sample, we focus on a single A+B pattern channel and use it to illustrate the process of attention mechanism. The corresponding patch embedding for one channel is in shape \( \mathbf{P}_d \in \mathbb{R}^{p \times D} \). To model temporal dependencies for each patch within each A+B pattern channel, we apply multi-head self-attention (MHSA) to each of the patches. For each head \( j \in \{1, \dots, h\} \), the output is computed as:
\begin{equation}
\text{SA}_j(\mathbf{P}_d) = \text{softmax} \left( \frac{(\mathbf{P}_d \mathbf{W}_j^Q)(\mathbf{P}_d \mathbf{W}_j^K)^\top}{\sqrt{d_k}} \right) (\mathbf{P}_d \mathbf{W}_j^V),\quad {MultiHead}(\mathbf{P}_d) = \big\|_{j=1}^{h} \text{SA}_j(\mathbf{P}_d) \cdot \mathbf{W}^O,
\end{equation}
where \( \mathbf{W}_j^Q, \mathbf{W}_j^K, \mathbf{W}_j^V \in \mathbb{R}^{D \times d_k} \) are the learnable weight matrices for the \( j \)-th head, \( \mathbf{W}^O \in \mathbb{R}^{h d_k \times D} \) is the final output projection, and \( d_k = D / h \) is the dimension per head. This way, the attention mechanism assigns adaptive weights to every patch pair, enabling the model to learn long-range temporal dependencies across the patches. By computing \( h \) attention heads in parallel, the model captures diverse features over the temporal structure of the input.

\noindent
\underline{\textit{Squeeze-and-Excitation Feature Recalibration}}:  
After the MHSA module outputs patch-wise encoded features \( \mathbf{P}_d \in \mathbb{R}^{p \times D} \) for each A+B pattern channel, we aim to recalibrate feature importance from a global perspective by assigning adaptive weights to each feature dimension within every channel. Returning to the macro view, the full output tensor can be expressed as \( \mathbf{X}' \in \mathbb{R}^{B \times d \times p \times D} \) after we established CVT-Conv layer, where \( B \) is the batch size, \( d \) is the number of A+B pattern channels, \( p \) is the number of patches, and \( D \) is the embedding dimension.

To achieve this, we apply a Squeeze-and-Excitation (SE) block. We first compute a compact summary \( \mathbf{g} \in \mathbb{R}^{B \times d} \) for each channel by performing global average pooling over the patch \( p \) and feature dimensions \( D \). The resulting summary vector is then passed through a two-layer fully connected network with ReLU and sigmoid activations to generate a set of soft importance weights \( \mathbf{s} \in \mathbb{R}^{B \times d} \). These weights are used to recalibrate the original tensor by broadcasting \( \mathbf{s} \) across the patch and feature dimensions via unsqueeze operations and applying element-wise multiplication. The complete process is defined as:
\begin{equation}
\begin{aligned}
&g_{b,d} = \frac{1}{p \cdot D} \sum_{a=1}^{p} \sum_{c=1}^{D} \mathbf{X}_{b,d,a,c}\ \ , \quad
\mathbf{s} = \sigma\left(\mathbf{W}_2 \cdot \text{ReLU}(\mathbf{W}_1 \mathbf{g})\right),\quad \mathbf{X}_{\text{SE}} = \mathbf{X'} \cdot \mathbf{s}.unsqueeze(-1).unsqueeze(-1)
\end{aligned}
\end{equation}
Here, \( \mathbf{W}_1 \in \mathbb{R}^{d \times d_r} \), \( \mathbf{W}_2 \in \mathbb{R}^{d_r \times d} \) are learnable weights of the excitation network, and \( \sigma(\cdot) \) denotes the sigmoid activation function. The recalibrated output \( \mathbf{X}_{\text{SE}} \in \mathbb{R}^{B \times d \times p \times D} \) preserves the original shape but with features of each channel scaled according to their learned importance. This mechanism allows the model to adaptively focus on seizure-relevant channels while suppressing less informative ones, improving both performance and interpretability.

\noindent
\underline{\textit{Seizure Risk Prediction}}:  
After recalibration, the output tensor \( \mathbf{X}_{\text{SE}} \in \mathbb{R}^{B \times d \times p \times D} \) is flattened across the channel, patch, and feature dimensions to form a global representation \( \mathbf{H} \in \mathbb{R}^{B \times (d \cdot p \cdot D)} \). This representation is passed through a linear layer followed by a dropout and sigmoid activation to produce the final seizure risk prediction:
\begin{equation}
\hat{Y} = \sigma(\mathbf{W} \mathbf{H} + \mathbf{b}), \quad \mathbf{W} \in \mathbb{R}^{(d \cdot p \cdot D) \times 1}, \quad \hat{Y} \in \mathbb{R}^{B \times 1}.
\end{equation}
Here, \( \mathbf{W} \) and \( \mathbf{b} \) are learnable parameters, and \( \sigma(\cdot) \) denotes the sigmoid function that transforms the output into a probability score. This step enables the model to aggregate global information across all channels and time patches to produce a sample-level prediction of seizure risk.

\noindent
\underline{\textit{Loss Function and Optimization}}:  
Since seizure risk forecasting is a binary classification task, the model is optimized using binary cross-entropy loss:
\begin{equation}
\mathcal{L} = -\frac{1}{M} \sum_{i=1}^{M} \left( y_i \log \hat{Y}_i + (1 - y_i) \log (1 - \hat{Y}_i) \right),
\end{equation}
\noindent
where \( y_i \) represents the ground truth label, taking values of either 0 (low-risk) or 1 (high-risk), while \( \hat{Y}_i \) denotes the model’s predicted probability of seizure occurrence. The loss is averaged over the batch size \( M \) to ensure stable optimization. Given the inherent class imbalance in seizure data, with significantly more low-risk days than high-risk days, binary cross-entropy mitigates bias toward the majority class by weighting the contribution of each prediction.

%% file: section/experi.tex
\section*{Experimental Settings}
\vspace{-3mm}

\underline{\textit{Datasets}}: Our study uses electrocorticographic data from patients with implanted RNS devices. Rather than raw spectrograms, we extract IEA surrogate biomarkers as features and LEs as seizure risk indicators. The dataset comprises recordings from five patients, each identified by a unique RNS Patient ID, and collected at the Emory Epilepsy Center. The daily recordings span from 3{,}030 to 6{,}953 days per patient. These patients were selected based on the availability of ultra-long-term monitoring data and clinically validated seizure annotations. The dataset is partitioned into 70\% training, 10\% validation, and 20\% testing sets.


\underline{\textit{Data Preprocessing and Labeling}}: 
To account for clinical variability and device reprogramming, A+B patterns are Z-score normalized per patient:
$
Z_t = \frac{X_t - \mu}{\sigma},
$
where \(X_t\) is the A+B count at time \(t\), and \(\mu, \sigma\) are the mean and standard deviation across all visits. For labeling, high-risk days (\(Y_t = 1\)) are identified using a dynamic, patient-specific threshold: days where LE count exceeds 70\% of the mean over the past 60 days. This adaptive strategy, informed by clinician input, reflects evolving seizure patterns and maintains robustness across clinical stages.

\begin{figure}[htbp]
    \centering
    \includegraphics[width=0.9\textwidth]{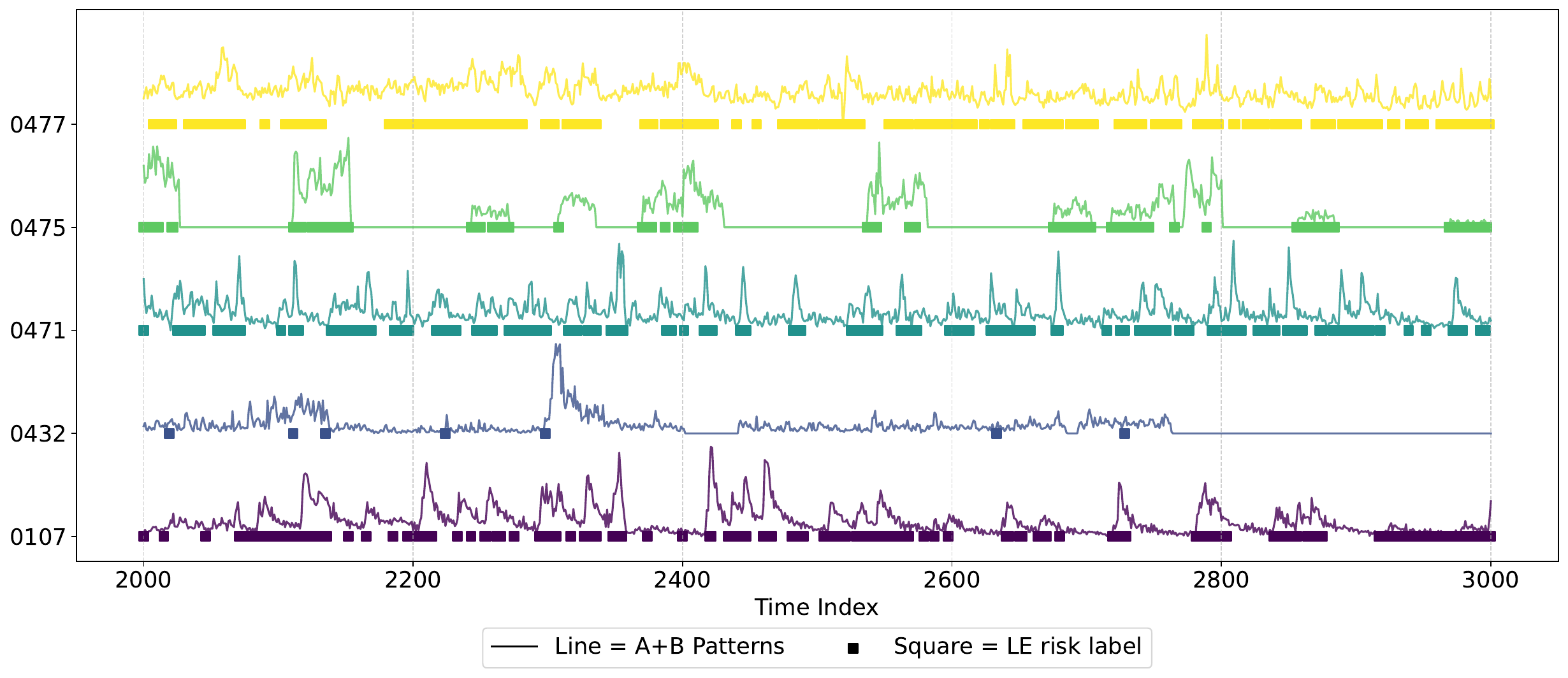}
    \caption{A+B patterns and LE risk labels over a 1000-day period. The line represents normalized A+B patterns, while the square markers indicate high risk labels.}
    \label{fig:channel_risk_1000}
\end{figure}

\underline{\textit{Data Visulization}}: To illustrate both macro-level trends and finer variability, we visualized 1000-day segments of normalized A+B patterns alongside corresponding LE risk labels. These samples were selected for their size and for representing diverse seizure risk profiles across patients. Figure~\ref{fig:channel_risk_1000} shows clear differences emerge across patients: \textit{Patient 0475} and \textit{Patient 0432} show periodic high-risk patterns—short cycles for 0475 and long, irregular ones for 0432—affecting predictability. \textit{Patient 0471} exhibits short, frequent cycles with high A+B variability, indicating unstable risk dynamics. In contrast, \textit{Patients 0477} and \textit{0107} show prolonged high-risk periods and noisy A+B patterns, suggesting a persistent ictal-interictal state and increased forecasting difficulty.



\underline{\textit{Evaluation Metrics}}: 
According to the label distribution shown in Figure~\ref{fig:channel_risk_1000}, the label distribution for all patients is imbalanced. Thus we use Area Under the Receiver Operating Characteristic curve (ROC AUC) and Area Under Precision/Recall curve (PR AUC) as the metrics.~\cite{Saito2015}.

\underline{\textit{Implementation Details}}:
Models undergo extensive hyperparameter tuning. We set the learning rate to 0.003, hidden dimension (\( D \)) to 128, batch size to 2048, and weight decay to 0.0001. The model uses 3 input channels (\textit{enc\_in} = 3), 2 decoder input channels (\textit{dec\_in} = 2), and 3 output channels (\textit{c\_out} = 3). The Transformer encoder comprises 3 layers with 2 attention heads and feedforward dimension 1024. Dropout is set to 0.2 to prevent overfitting. To address class imbalance, we apply class weighting in the loss function using a \textit{pos\_weight} derived from the ratio of negative to positive samples per patient. Experiments are conducted on an NVIDIA TITAN RTX (24GB VRAM) with CUDA 12.4. Models are trained for up to 30 epochs with early stopping (patience = 5), based on validation AUC.

\underline{\textit{Baselines}}:
We compare SeizureFormer with a comprehensive set of baselines across three categories: statistical, machine learning, and deep learning. Statistical models, including Generalized Linear Models (GLM) and Poisson Regression \cite{nelder1972generalized, hilbe2011logistic}, provide interpretable baselines suited for structured medical data. For classical machine learning, we test Support Vector Machines (SVM) \cite{cortes1995support} and Logistic Regression \cite{hilbe2011logistic}, which are effective for binary classification using structured features. Deep learning models include GRU-LSTM \cite{cho2014learning} for capturing long-range dependencies, DLinear \cite{zeng2023dlinear} for decomposing trend and seasonality in time series, and PatchTST \cite{Nie2022PatchTST}, a Transformer-based model that leverages patching and global attention for improved forecasting.






%% file: section/result.tex
\section*{Experimental Results}  
In this section, we present our experimental results, structured around four key research questions (RQs) to systematically evaluate the performance of our proposed model, SeizureFormer, in seizure risk forecasting.  

\begin{compactenum}[\textbullet]
    \item \textbf{RQ1: Does SeizureFormer outperform baseline models?}  
    \item \textbf{RQ2: How does the prediction window (\textit{PredLen} = 1, 3, 7, 14 days) affect forecasting performance?}  
    \item \textbf{RQ3: How do different patients’ ictal patterns influence model performance?} 
    \item \textbf{RQ4: How do different components affect the performance of SeizureFormer?} 
\end{compactenum}  

\begin{table}[htbp!]
\centering
\caption{Seizure Risk Forecasting Results. The table presents each models' seizure risk forecasting performance under different settings. Mean ROC AUC and Mean PR AUC are also calculated to show overall performance of each model across different settings. The best performance is bold and the second best performance is underlined.}
\label{tab:results}
\resizebox{0.95\textwidth}{!}{%
\begin{tabular}{cc|cc|cc|cc|cc|cc|cc}
\toprule
\multicolumn{2}{c|}{Patients} & \multicolumn{2}{c|}{0107} & \multicolumn{2}{c|}{0432} & \multicolumn{2}{c|}{0471} & \multicolumn{2}{c|}{0475} & \multicolumn{2}{c|}{0477} & \multicolumn{2}{c}{Average} \\
Models & Pred Lengths & ROC AUC & PR AUC & ROC AUC & PR AUC & ROC AUC & PR AUC & ROC AUC & PR AUC & ROC AUC & PR AUC & ROC AUC & PR AUC \\
\midrule
\multirow{4}{*}{GLM}&
1  & 59.5\% & 37.6\% & 77.8\% & 19.6\% & 68.9\% & 55.9\% & 95.9\% & 88.7\% & 81.2\% & 67.1\% & \multirow{4}{*}{71.32\%} & \multirow{4}{*}{64.34\%} \\
& 3  & 55.2\% & 51.5\% & 69.8\% & 21.8\% & 67.6\% & 76.3\% & 92.5\% & 86.8\% & 76.8\% & 75.8\% & & \\
& 7  & 58.7\% & 70.2\% & 53.4\% & 25.9\% & 67.5\% & 87.3\% & 90.2\% & 87.0\% & 80.4\% & 88.4\% & & \\
& 14 & 58.2\% & 81.7\% & 21.5\% & 69.4\% & 95.3\% & 86.4\% & 84.1\% & NA & NA & NA & & \\

\midrule
\multirow{4}{*}{Poisson Regression}&
1  & 62.7\% & 48.8\% & 78.0\% & 27.1\% & 67.7\% & 54.7\% & 96.2\% & 90.8\% & 80.8\% & 65.9\% & \multirow{4}{*}{71.20\%} & \multirow{4}{*}{65.24\%} \\
& 3  & 54.6\% & 51.3\% & 78.8\% & 33.5\% & 63.0\% & 73.6\% & 93.6\% & 88.5\% & 71.5\% & 69.4\% & & \\
& 7  & 59.8\% & 70.6\% & 72.1\% & 31.2\% & 59.6\% & 85.7\% & 90.5\% & 87.6\% & 76.9\% & 87.2\% & & \\
& 14 & 65.0\% & 82.8\% & 38.2\% & 17.2\% & 61.6\% & 93.7\% & 82.2\% & 79.9\% & NA & NA & & \\

\midrule\midrule
\multirow{4}{*}{Logistic Regression}&
1  & 62.4\% & 44.9\% & 82.3\% & 22.3\% & 68.4\% & 54.8\% & 94.8\% & 87.1\% & 79.3\% & 62.6\% & \multirow{4}{*}{73.13\%} & \multirow{4}{*}{64.94\%} \\
& 3  & 53.9\% & 49.1\% & 77.9\% & 28.2\% & 65.3\% & 73.8\% & 93.2\% & 87.6\% & 76.6\% & 74.6\% & & \\
& 7  & 55.3\% & 66.7\% & 70.9\% & 31.7\% & 68.7\% & 88.1\% & 90.4\% & 87.4\% & 81.3\% & 89.2\% & & \\
& 14 & 59.1\% & 82.2\% & 54.3\% & 25.3\% & 69.8\% & 95.3\% & 85.6\% & 82.9\% & NA & NA & & \\

\midrule
\multirow{4}{*}{SVM}&
1  & 60.4\% & 39.4\% & 64.4\% & 16.4\% & 68.8\% & 55.2\% & 95.8\% & 88.5\% & 79.3\% & 63.0\% & \multirow{4}{*}{71.35\%} & \multirow{4}{*}{63.92\%} \\
& 3  & 55.5\% & 51.3\% & 72.9\% & 22.9\% & 68.1\% & 76.1\% & 93.3\% & 87.9\% & 75.5\% & 71.8\% & & \\
& 7  & 57.3\% & 68.4\% & 62.1\% & 27.0\% & 67.5\% & 88.0\% & 90.6\% & 87.4\% & 81.3\% & 89.2\% & & \\
& 14 & 57.7\% & 81.4\% & 50.8\% & 22.9\% & 69.0\% & 95.2\% & 85.4\% & 82.4\% & NA & NA & & \\

\midrule\midrule
\multirow{4}{*}{DLinear}&
1  & \underline{70.9\%} & 57.7\% & 46.2\% & 8.96\% & 68.9\% & 55.9\% & 96.9\% & 92.0\% & 81.8\% & 67.6\% & \multirow{4}{*}{72.83\%} & \multirow{4}{*}{65.97\%} \\
& 3  & 52.9\% & 52.9\% & 73.4\% & 22.7\% & 69.4\% & 78.0\% & 93.6\% & 88.9\% & 79.3\% & 78.6\% & & \\
& 7  & 47.1\% & 61.6\% & 84.2\% & 37.2\% & 69.1\% & 89.5\% & 88.2\% & 85.1\% & 79.0\% & \underline{89.6\%} & & \\
& 14 & 52.5\% & 77.5\% & 80.6\% & 35.6\% & 69.5\% & 95.2\% & 80.3\% & 78.9\% & NA & NA & & \\

\midrule
\multirow{4}{*}{GRU\_LSTM}&
1  & 58.2\% & \underline{81.7\%} & 82.5\% & 32.4\% & 68.5\% & 56.3\% & {\textbf{97.6\%}} & \underline{93.3\%} & \underline{83.9\%} & 68.4\% & \multirow{4}{*}{73.16\%} & \multirow{4}{*}{71.05\%} \\
& 3  & 67.9\% & 66.2\% & 87.8\% & 54.3\% & 70.8\% & 78.7\% & 94.7\% & 88.9\% & 76.7\% & 73.5\% & & \\
& 7  & 64.5\% & 75.1\% & 76.8\% & 30.8\% & 58.4\% & 88.1\% & 90.2\% & 86.2\% & 50.0\% & 85.2\% & & \\
& 14 & 45.6\% & 77.6\% & 35.1\% & 51.9\% & 51.9\% & 85.7\% & 82.6\% & NA & NA & NA & & \\

\midrule
\multirow{4}{*}{PatchTST}&
1  & \textbf{74.2\%} & 63.9\% & \underline{93.2\%} & 62.4\% & 66.4\% & 46.0\% & \underline{97.1\%} & 91.7\% & 78.2\% & 62.7\% & \multirow{4}{*}{\underline{76.41\%}} & \multirow{4}{*}{\underline{73.07\%}} \\
& 3  & 52.0\% & 45.4\% & 92.6\% & 59.6\% & 69.6\% & 76.4\% & 94.5\% & 89.9\% & 77.4\% & 78.7\% & & \\
& 7  & 53.7\% & 70.8\% & 87.4\% & 49.5\% & 73.1\% & 91.0\% & 91.2\% & 88.0\% & 78.8\% & 89.2\% & & \\
& 14 & 28.9\% & 64.3\% & 83.9\% & {\textbf{78.5\%}} & {\textbf{74.6\%}} & {\textbf{95.9\%}} & 85.0\% & 84.5\% & NA & NA & & \\

\midrule\midrule
\multirow{4}{*}{\textbf{SeizureFormer}}&
1  & \textbf{74.2\%} & 53.5\% & {\textbf{94.7\%}} & 61.6\% & 70.3\% & 59.2\% & \underline{97.1\%} & {\textbf{93.4\%}} & {\textbf{84.8\%}} & 73.0\% & \multirow{4}{*}{{\textbf{79.44\%}}} & \multirow{4}{*}{{\textbf{76.29\%}}} \\
& 3  & 61.4\% & 68.1\% & 92.2\% & \underline{77.8\%} & 71.0\% & 76.7\% & 94.8\% & 90.8\% & 80.6\% & 80.7\% & & \\
& 7  & 47.7\% & 64.6\% & 87.9\% & 52.3\% & 71.8\% & 90.6\% & 92.7\% & 88.3\% & 81.6\% & {\textbf{90.7\%}} & & \\
& 14 & 64.7\% & \textbf{83.1\%} & 85.5\% & 64.6\% & \underline{73.3\%} & \underline{95.7\%} & 86.0\% & 84.9\% & NA & NA & & \\
\bottomrule
\end{tabular}%
}
\end{table}



\underline{\textit{RQ1 - Our Model's Performance}}:
To evaluate whether SeizureFormer outperforms baseline models, we first analyze the overall model performance, disregarding the effects of prediction length (\textit{PredLen}) and patient-specific variations. Table~\ref{tab:results} presents the ROC AUC and PR AUC scores across all patients and \textit{PredLen} values. SeizureFormer achieves the highest average ROC AUC (\textit{79.44\%}) and PR AUC (\textit{76.29\%}), surpassing all baseline models. The second-best performing model, PatchTST, achieves \textit{76.41\%} ROC AUC and \textit{73.07\%} PR AUC, while other deep learning models, such as GRU-LSTM, also demonstrate strong performance but fall short of SeizureFormer. Traditional statistical and classical machine learning models, including Logistic Regression, Poisson Regression, GLM, and SVM, yield lower mean PR AUC scores, reflecting their limited capacity to model complex temporal dynamics and address severe class imbalance.
These results establish SeizureFormer as the most effective model overall, though further analysis is needed to assess its robustness across different prediction lengths and patient cases.
Examining patient-specific performance, SeizureFormer achieves top performance for four out of five patients, demonstrating strong generalization across different seizure patterns. For Patient \textit{0477}, where long-term forecasting becomes challenging due to data collapse at \textit{PredLen} = \textit{14}, SeizureFormer still outperforms other models in shorter forecasts. GRU-LSTM exhibit strong performance in some cases but struggle with consistency, particularly for longer \textit{PredLen}. These results confirm SeizureFormer as the most effective model for seizure risk forecasting.


\underline{\textit{RQ2 - Impact of prediction window}}:  
To analyze the impact of prediction window length on model performance, we evaluate trends in both ROC AUC and PR AUC scores across different \textit{PredLen} values. Table~\ref{tab:results} provides the full performance breakdown.
A general pattern emerges where shorter \textit{PredLen} values (\textit{1, 3} days) yield higher ROC AUC scores, while longer \textit{PredLen} values (\textit{7, 14} days) lead to higher PR AUC scores.  
For \textit{PredLen} = \textit{1, 3} days, models show higher ROC AUC, with deep learning models (SeizureFormer, PatchTST, GRU-LSTM) performing consistently well. Traditional models (e.g., Poisson Regression, SVM) show weaker and less stable performance. For \textit{PredLen} = \textit{7, 14} days, PR AUC improves, reflecting better long-term risk trend capture. Deep learning models handle class imbalance more effectively, with SeizureFormer exceeding \textit{80\%} PR AUC and GRU-LSTM often above \textit{70\%}. Statistical models frequently fall below \textit{50\%}. The trade-off between ROC AUC and PR AUC highlights the balance between short-term precision and long-term generalization. SeizureFormer performs robustly across all \textit{PredLen} values.

\underline{\textit{RQ3 - Variation between Patients}}:  
To investigate the impact of different patients' ictal patterns on model performance, we analyze how seizure forecasting varies across individuals. Given the variability in seizure occurrence and EEG characteristics, it is essential to assess whether certain models maintain consistent performance across different patients.
Table~\ref{tab:results} shows that forecasting performance varies widely across patients. Some (e.g., \textit{0475}) consistently achieve high scores, while others (e.g., \textit{0432}) show large fluctuations, indicating varying signal clarity and model sensitivity.
SeizureFormer performs most consistently, while GRU-LSTM shows more variation. Statistical models often fail on patients with complex patterns.
Model performance is strongly patient-dependent, with SeizureFormer generalizing better than other baselines.

\begin{table}[htbp]
\centering
\caption{Ablation experiment results on Patient 0477. The best performance is bold and the second best is underlined.}
\label{tab:ablation}
\resizebox{\textwidth}{!}{%
\begin{tabular}{c|cc|cc|cc}
\toprule
\textbf{Model Variant} & \multicolumn{2}{c|}{\textbf{PredLen = 1}} & \multicolumn{2}{c|}{\textbf{PredLen = 3}} & \multicolumn{2}{c}{\textbf{PredLen = 7}} \\
\midrule
\textbf{Metrics} & ROC AUC & PR AUC & ROC AUC & PR AUC & ROC AUC & PR AUC \\
\midrule
\textbf{Full Model} & \textbf{84.83\%} & \textbf{73.03\%} & \textbf{80.62\%} & \textbf{80.70\%} & \underline{81.63\%} & \textbf{90.70\%} \\
w/o CNN Patch Embedding & \underline{84.39\%} & \underline{71.89\%} & 73.31\% & 75.60\% & 64.61\% & 81.59\% \\
w/o SE Block & 82.80\% & 68.68\% & \underline{77.95\%} & \underline{78.21\%} & \textbf{82.76\%} & \underline{90.20\%} \\
w/o Cross-Variable Temporal convolution & 77.71\% & 63.55\% & 77.66\% & 77.24\% & 75.69\% & 88.65\% \\
w/o All Modules & 76.35\% & 61.86\% & 70.87\% & 72.40\% & 72.97\% & 87.56\% \\
\bottomrule
\end{tabular}%
}
\end{table}

\underline{\textit{RQ4 - Each Module's Impact on SeizureFormer}}: To evaluate the role of individual components in \textit{SeizureFormer}, we conducted ablation experiments on \textbf{Patient 0477}, a representative case with persistent high-risk dynamics. We tested the effect of removing three core modules: the CNN-based multi-kernel patch embedding, the cross-variable temporal 2D convolution (CVT-Conv), and the squeeze-and-excitation (SE) block, as well as all modules together. Results under different prediction lengths (1, 3, and 7 days) are presented in Table~\ref{tab:ablation}.
Removing any module reduces performance. The CNN-based patch embedding has the largest impact, especially at \textit{PredLen = 7}, with PR AUC dropping from 90.70\% to 81.59\%, showing its importance for long-range features. Removing the cross-variable 2D convolution also consistently lowers performance, particularly at \textit{PredLen = 1} and \textit{3}, indicating its role in short-term precision.
At \textit{PredLen = 7}, removing the SE block slightly improves ROC AUC but lowers PR AUC, which is more clinically meaningful. Its removal also harms performance at shorter horizons, emphasizing its value for early precision.
These results highlight the complementary roles of all components in achieving robust and well-calibrated forecasting.



%% file: section/conclusion.tex
\section*{Conclusion} 
\vspace{-3mm}

We present SeizureFormer, a Transformer-based model that advances seizure risk forecasting using RNS-extracted IEA surrogated biomarkers. By combining muti-kernel CNNs, cross-channel convolution, self-attention, and Squeeze \& excitation module, it captures both short- and long-term temporal patterns. Experiments show state-of-the-art performance and strong generalizability across patients and prediction windows, paving the way for proactive, personalized epilepsy care. Future directions include transfer learning, adaptive biomarker modeling, and integration with closed-loop neurostimulation.